# Inferring trust in recommendation systems from brain, behavioural, and physiological data

Vincent K.M. Cheung [a*], Pei-Cheng Shih [a, b], Masato Hirano [a, b], Masataka Goto [c], Shinichi Furuya [a, b]

[a] Sony Computer Science Laboratories, Inc., 3-14-13, Higashigotanda, Shinagawa-ku, Tokyo, 141-0022 Japan

[b] NeuroPiano Institute, 13-1, Hontoro-cho, Shimogyo, Kyoto, 600-8068 Japan

[c] National Institute of Advanced Industrial Science and Technology (AIST), 1-1-1 Umezono, Tsukuba, Ibaraki, 305-8560 Japan

*Correspondence:
cheung@csl.sony.co.jp
Sony Computer Science Laboratories, Inc., 3-14-13, Higashigotanda, Shinagawa-ku, Tokyo, 141-0022 Japan

## Acknowledgements
This study was supported by JST CREST (Grant number JPMJCR20D4).

## Declaration of Interest
No competing interests to declare

## Author contributions
Conceptualization: VKMC, PCS, MH, SF; Data curation: VKMC; Formal analysis: VKMC, PCS, MH; Funding acquisition: SF, MG; Investigation: VKMC; Methodology: VKMC, PCS, MH; Project administration: SF, MG; Resources: SF, MG; Software: VKMC, PCS, MH; Supervision: SF, MG; Validation: VKMC; Visualization: VKMC; Writing – original draft: VKMC; Writing – review & editing: VKMC, MG, SF, MH, PCS

## Abstract

As people nowadays increasingly rely on artificial intelligence (AI) to curate information and make decisions, assigning the appropriate amount of trust in automated intelligent systems has become ever more important. However, current measurements of trust in automation still largely rely on self-reports that are subjective and disruptive to the user. Here, we take music recommendation as a model to investigate the neural and cognitive processes underlying trust in automation. We observed that system accuracy was directly related to users' trust and modulated the influence of recommendation cues on music preference. Modelling users' reward encoding process with a reinforcement learning model further revealed that system accuracy, expected reward, and prediction error were related to oscillatory neural activity recorded via EEG and changes in pupil diameter. Our results provide a neurally grounded account of calibrating trust in automation and highlight the promises of a multimodal approach towards developing trustable AI systems.

# Introduction

The rise and permeation of automation—particularly artificial intelligence (AI)—has promised increased productivity, connectivity, and individuality [1]. As our lives become increasingly entwined and integrated with automation, the question of how much trust people put in automated systems has never been more important [2].

In the context of automation, a useful definition of trust is the "attitude that an agent will help achieve an individual's goals in a situation characterized by uncertainty and vulnerability" [3]. While trusting another person involves assessing their *ability* in carrying out the task, *benevolence* in having the trustor's best interests, and *integrity* of sharing similar principles [4], trusting an automated system requires examining its *performance*, *purpose*, and *process* [3]: Performance is about how well the system does its job. Process is about the algorithms a system uses to accomplish its tasks. Purpose is about understanding the intentions behind why the system was built. On one hand, mistrusting, or being overly reliant and complacent on automation, often leads to failure in identifying errors or the inability to handle them, and could cause unexpected outcomes or even accidents. On the other hand, distrusting, or the unwillingness to engage an automated system could lead to task inefficiencies. Assigning an appropriate, or *calibrated*, amount of trust is therefore crucial to the effective and safe use of automation [3,5].

Trust in automation is primarily assessed via user self-reports. Beyond free responses, many psychometric measures exist with the aim to measure trust in a structured and standardised manner (see [6] for review). Two measures stand out in particular: the Checklist for Trust between People and Automation [7], a 12-item Likert scale questionnaire covering reliability, integrity, and overall trust, as well as the Trust Perception Scale for Human Robot Interaction [8] that is specialised for human-robot interaction.

While self-reports can provide a comprehensive view of users' trust, they are however limited [6]. First, self-reports necessitate user engagement and their explicit recognition of trust, which may be problematic if users give biased and uncalibrated responses. Second, self-reports cannot be continuously assessed and could be disruptive if queried too often. Third, self-reports provide a conceptual account of trust, but do not explain the underlying computational mechanisms that give rise to such a concept. These call for implicit measures of user trust.

To that end, several studies have explored the neurophysiological correlates of trust in automation by examining differences in evoked activity when observing errors committed by automated agents. While observing errors typically elicit an observational error-related negativity (oERN) and observational error positivity (oPe) that is similar but possibly distinct from the N2-P3 complex (a negativity at around 200 ms and subsequent positivity at around 300 ms recorded via electroencephalography (EEG)) [9], the error positivity has shown to be reduced when a less-trusted [10] or less-reliable [11] automated agent performed a modified flanker task. The observation error-evoked activity has been argued to be related to effective connectivity from the dorsal anterior cingulate cortex (ACC) to the posterior cingulate cortex (PCC) [12], which have also been implicated in reciprocal trust [13] and deliberation on whether to utilise agent advice [14]. Studies aiming to decode users' trust in automated agents from EEG data further identified oscillatory activity in frontal electrodes as relevant features encoding trust [15–18].

In addition, pupil activity has been shown to encode social trust. For example, [19] found that trust in an in-group partner was facilitated by the mimicry of pupil size and was moderated by oxytocin. Using functional magnetic resonance imaging (fMRI), pupil mimicry was identified to be related to activation of the Theory-of-Mind network, including the precuneus, temporoparietal junction, superior temporal sulcus, and medial prefrontal cortex [20]. Pupil dilation has furthermore been shown to be related to positive traits such as increased trust, friendliness, and attractiveness, as well as increased reciprocity [21].

While these studies provide an initial foray into the neurocognitive basis of trust in automation, it is surprising that this has not yet been explored in an important class of automated systems—recommendation systems—which are already well-integrated into the lives of many in the world with the prevalence of streaming platforms [22]. Unlike the use of automated agents in specialised tasks (e.g. modified flanker task, reciprocal trust game) in previous studies, examining the neural computational mechanisms underlying trust in recommendation systems ensures that results are ecologically valid and generalisable to a wide population.

Here, we adapted a classical reward prediction paradigm to test how users' trust in a music recommendation system influenced their enjoyment when listening to its recommended songs, as well as investigated its neurophysiological correlates (see Figure 1 and Methods). Thirty subjects were presented with three music recommendation systems that varied in accuracy whilst their pupil and oscillatory neural activity were recorded. At the start of each trial, the recommendation system presented a like or dislike cue to the ensuing song with equal likelihood, whose valence has been previously rated by the subject. Following song presentation (10 s), subjects rated their preference on a 1–9 scale, as well as how much they trusted the recommendation system on a 1–9 scale at every $10^{th}$ trial. In short, we found that system accuracy predicted users' trust, differences in theta-band activity, as well as modulated increases in pupil size due to mismatches in expected reward.

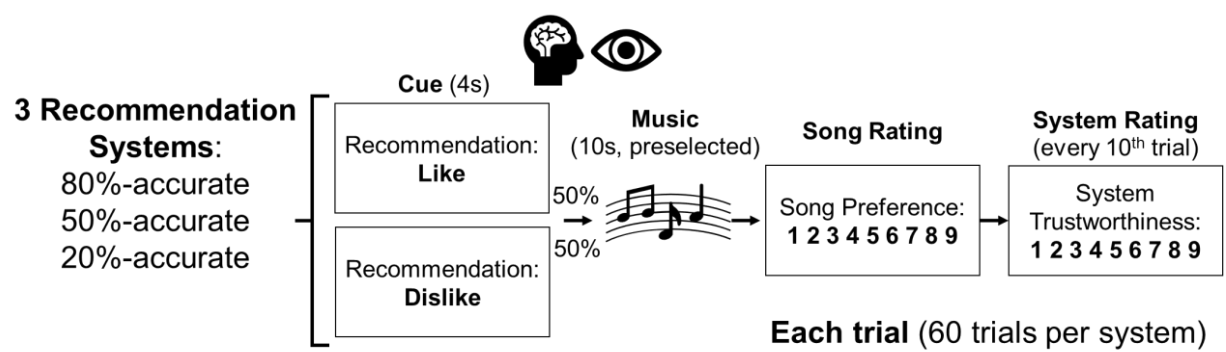

Figure 1. Experimental paradigm. Subjects were exposed to three music recommendation systems with varying accuracies. At each trial, the recommendation system provided a cue that the ensuing music excerpt would be liked or disliked by the subject with equal probability. After music presentation, subjects rated their song preference on a 1-9 scale, as well as their trust on the system every 10 trials. Subjects' neural and physiological activity were recorded via electroencephalography (EEG) and pupillometry throughout the experiment.

# Results

*Recommendation cues modify users' song preference, depending on trust in the system*

First, we tested subjects' rating on how trustable a recommendation was depended on system accuracy over time (Figure 2). Resolving the significant main effect of accuracy ($F(2, 27.0) = 34.1$, $p = 4.17 \times 10^{-8}$) revealed that, averaged across trials, the 80%-accurate system was rated as significantly more trustable than the 50%-accurate system ($t(26.9) = 5.47$, $p_{corrected} = 2.56 \times 10^{-5}$), and the 50%-accurate system as more trustable than the 20%-accurate system ($t(27.1) = 3.17$, $p_{corrected} = 0.0101$). The absence of a significant main effect of trial ($F(5, 365) = 0.684$, $p = 0.636$) and accuracy*trial interaction ($F(10, 383) = 1.39$, $p = 0.185$) further suggest that subjects were sensitive to the accuracy of the system already from the first 10 trials. These demonstrate that trust in a recommendation system is accuracy-dependent, rapidly established, and persistent over time.

Next, we examined whether system accuracy and recommendation cue affected subjects' preference ratings on songs that they already liked or disliked (Figure 3). As expected, a main effect of actual song preference ($F(1, 35.8) = 132$, $p = 1.50 \times 10^{-13}$) indicates that actually liked songs were rated significantly higher than disliked songs. Interestingly, a significant main effect of cue ($F(1, 35.8) = 132$, $p = 3.63 \times 10^{-7}$) and cue*actual interaction ($F(1, 4802) = 5.35$, $p = 0.0208$) demonstrate that the recommendation cue could in fact shape subjects' song ratings: resolving the interaction, we found that dislike cues lowered subjects' preference ratings to actual liked songs by –0.781 ($Z = –6.79$, $p = 1.11 \times 10^{-11}$), while like cues raised subjects' preference ratings to actual disliked songs by a smaller extent, namely by 0.547 ($Z = 4.76$, $p = 1.96 \times 10^{-6}$). We nevertheless did not detect any accuracy-related changes in song preference with this model (main effect: $F(2, 4715) = 1.26$, $p = 0.283$; accuracy*cue: $F(2, 4807) = 0.682$, $p = 0.506$; accuracy*actual: $F(2, 4715) = 0.153$, $p = 0.858$; accuracy*cue*actual: $F(2, 4806) = 0.662$, $p = 0.516$).

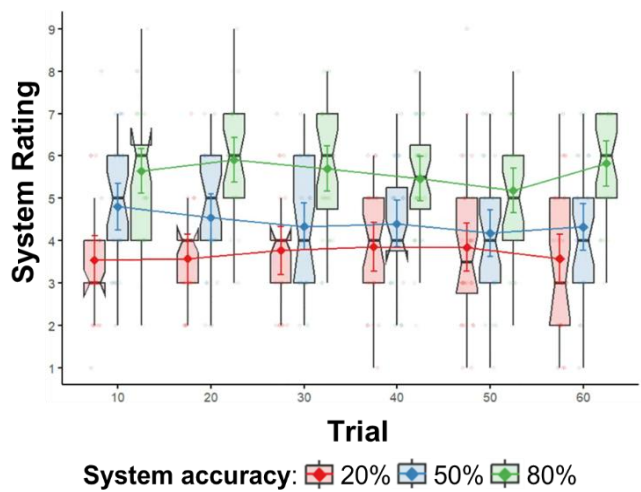

Figure 2. Trust in recommendation system as a function of system accuracy and trial. Subjects were sensitive to system accuracy already from the first 10 trials and remained consistent levels of trust throughout their exposure to the recommendation system.

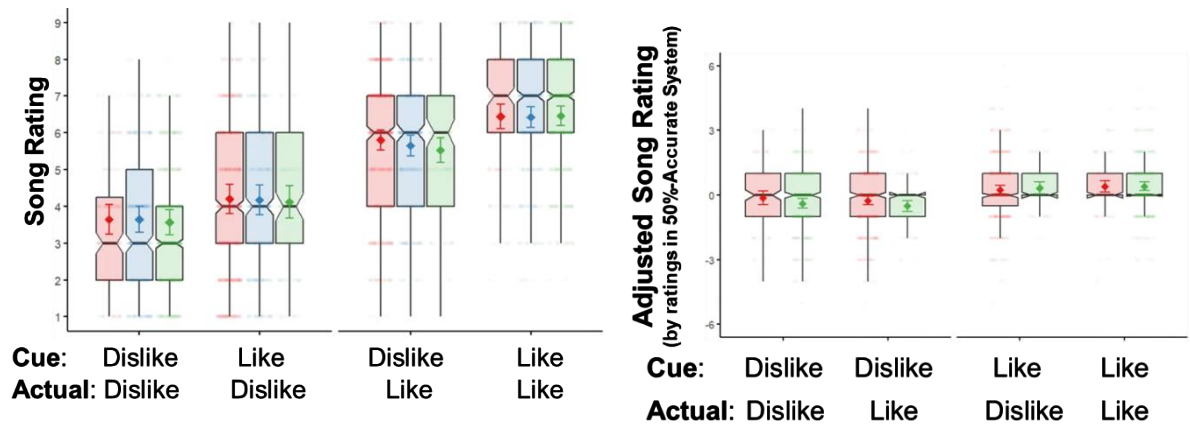

Figure 3. Recommendation cue influences song preferences, particularly when trusted. Across songs, subjects gave higher ratings for songs they identified as dislike if it was proceeded by a like cue, and conversely a lower rating for liked songs given a dislike cue (Left). When examining on a song-by-song level, the influence of cue was larger for dislike cues when the system was more trusted (Right).

While the previous analysis might suggest that system accuracy did not affect the influence of recommendation cue on subjects' song preference, it is plausible the effect was masked from averaging across stimuli. We therefore carried out a control analysis where the rating of each stimulus presented during the 80%- and 20%-accurate system was first normalised by its rating when presented during the 50%-accurate system (Figure 3). After normalisation, we indeed saw a significant accuracy*cue interaction ($F(1, 3184) = 5.77$, $p = 0.0164$): Resolving the interaction revealed that subjects gave significantly lower preference ratings (–0.255) to the same songs in the 80%-accurate system compared to 20%-accurate system only for dislike ($Z = 2.27$, $p = 0.0231$) but not like ($Z = 0.393$, $p = 0.694$) cues. This suggests that increased trust in a recommendation system enhanced the negative influence of dislike cues in song preference. Replicating the previous analysis, we also observed a significant main effect of cue ($F(1, 34.7) = 37.3$, $p = 5.77 \times 10^{-7}$) and marginally significant cue*actual interaction ($F(1, 3179) = 3.68$, $p = 0.0553$). No other significant effects were detected in this model (main effect of accuracy: $F(1, 38.2) = 1.27$, $p = 0.266$; main effect of actual like/dislike: $F(1, 71.2) = 0.00$, $p = 0.995$; accuracy*actual: $F(1, 3077) = 0.131$, $p = 0.717$; accuracy*cue*actual: $F(1, 3184) = 0.122$, $p = 0.727$).

*Computational modelling reveals that users' expected reward is accuracy-dependent*

We next simulated subjects' expected reward upon seeing like and dislike cues in the cue phase when exposed to the three recommendation systems using a reinforcement learning model. Examining the expected reward traces (Figure 4), we noticed a gradual increase in expected reward for like cues and decrease for dislike cues in the 80%-accurate system. This suggests that subjects learnt to associate like cues with positive reward, and dislike cues with negative reward. On the other hand, the expected reward hovered around the initial value for both like and dislike cues in the 50%-accurate system, likely because did not provide useful information. Strikingly, contrary to the 80%-accurate system, we found that expected reward tended to increase with dislike cues and decrease with like cues in the 20%-accurate system. Since subjects were never told the accuracy of each system, this demonstrates that subjects *implicitly* learnt to associate like cues with lower reward and dislike cues with higher reward.

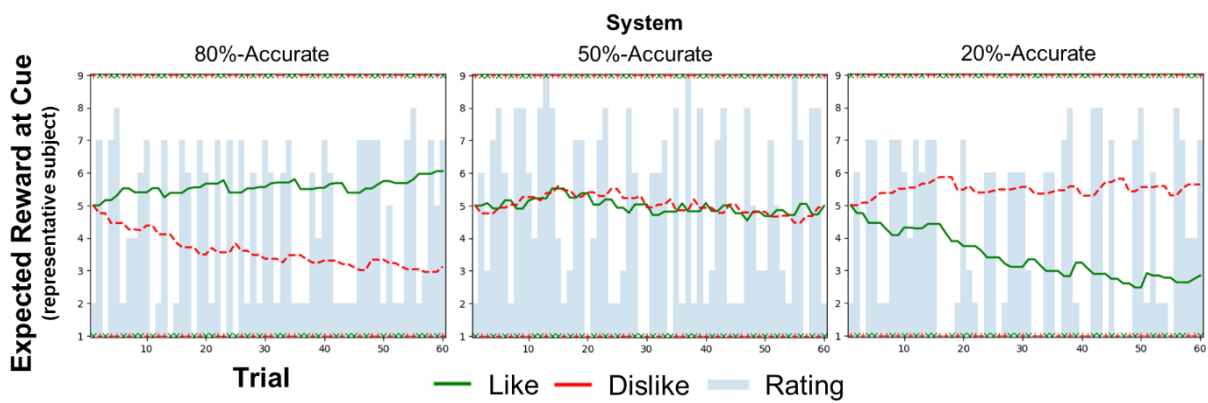

Figure 4. Expected reward during cue presentation across trials from a representative subject. Reward traces were derived from the Rescorla-Wagner reinforcement learning model. Crosses on top indicate actual cue presented at a given trial. Observe that like cues encoded a lower expected reward compared to dislike cues for the 20%-Accurate system.

To quantify the effect of system accuracy on expected reward, we computed the mean difference between expected reward upon seeing like versus dislike cues for each system across subjects (Figure 5). Echoing the previous observation, we saw a significant effect of accuracy ($F(2, 56) = 67.3$, $p = 1.28 \times 10^{-15}$), where the mean difference in expected reward between like and dislike cues was 1.30, 0.419, and –0.467 for the 80%-, 50%-, and 20%-accurate systems, respectively, and significantly differed across systems (80% vs 50%: $t(56) = 5.79$, $p_{corrected} = 9.92 \times 10^{-7}$; 80% vs 20%: $t(56) = 11.6$, $p_{corrected} = 7.32 \times 10^{-12}$; 50% vs 20%: $t(56) = 5.81$, $p_{corrected} = 9.17 \times 10^{-7}$). These show that subjects' expected reward depended on system accuracy.

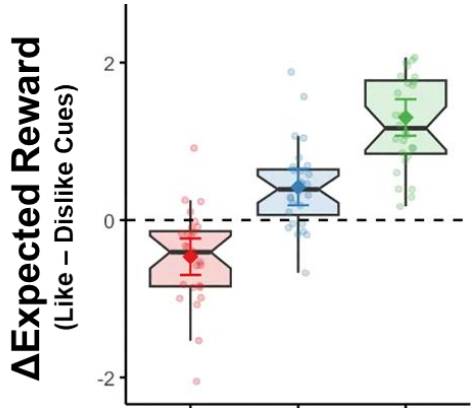

Figure 5. Mean trial-wise difference between expected reward of like and dislike cues differ significantly across system accuracy. Notably, 20%-accurate system exhibit expectation reversal, where expected reward was higher for dislike versus like cues.

*Predicting users' trust in a recommendation system from physiological and neural data*
We then examined the relationship between system accuracy, cue, and actual song preference on subjects' neural and pupil responses in the cue and music phases.

In the cue phase, we detected a significant main effect of system accuracy in subjects' event-related spectral perturbation (ERSP) at midfrontal EEG electrodes in the beta and theta bands. In both frequency bands, ERSP was highest in the 50%-, followed by the 20%- and 80%-accurate system across all trials (Figure 6 and Supplementary Table 1). The difference in ERSP was moreover significant between the 80%- and 50%-accurate systems for both frequency bands (beta: mean = 0.144, $Z = 2.42$, $p_{corrected} = 0.0415$; theta: mean = 0.449, $Z = 4.48$, $p_{corrected} = 2.24 \times 10^{-5}$), between 80%- and 20%-accurate systems in the theta band (mean = 0.362, $Z = 3.255$, $p_{corrected} = 0.00326$), but not between 20%- and 50%- systems in both bands (beta: mean = 0.139, $Z = 2.29$, $p_{corrected} = 0.0571$; theta: mean = 0.0862, $Z = 0.855$, $p_{corrected} = 0.669$) and between 80%- and 20%- systems in the beta band (mean = 0.00460, $Z = 0.0694$, $p_{corrected} = 0.997$). No other accuracy, cue, or actual preference-related effects were detected from ERSP. In addition, no accuracy, cue, or actual preference-related effects were detected from subjects' pupil diameter in the cue phase (Figure 6).

In the music phase, we likewise detected a significant main effect of system accuracy in the theta band (see Supplementary Table 2). As before, ERSP was highest across trials in the 50%-accurate system, followed by the 20%- and the 80%-accurate system (50% vs 20%: mean = 0.0459, $Z = 0.579$, $p_{corrected} = 0.832$; 50% vs 80%: mean = 0.280, $Z = 3.54$, $p_{corrected} = 0.00115$; 20% vs 80%: mean = 0.234, $Z = 2.67$, $p_{corrected} = 0.0208$). Furthermore, we saw a significant main effect of actual liked/disliked songs in the alpha band, with lower ERSP during presentation of liked compared to disliked songs (mean = 0.130, $Z = 1.99$, $p = 0.0464$). On the other hand, we also detected a significant accuracy*cue*actual interaction for pupil diameter (Figure 6). Resolving this interaction, we found that the effect was driven by an increase in pupil diameter for actual liked vs disliked songs following dislike cues in the 80%-accurate system (mean = 0.567, $Z = 2.42$, $p = 0.0156$), but a decrease in pupil diameter for actual liked vs disliked songs following dislike cues in the 20%-accurate system (mean = 0.483, $Z = 2.10$, $p = 0.0360$). No other accuracy, cue, or actual preference-related effects were detected in the music phase.

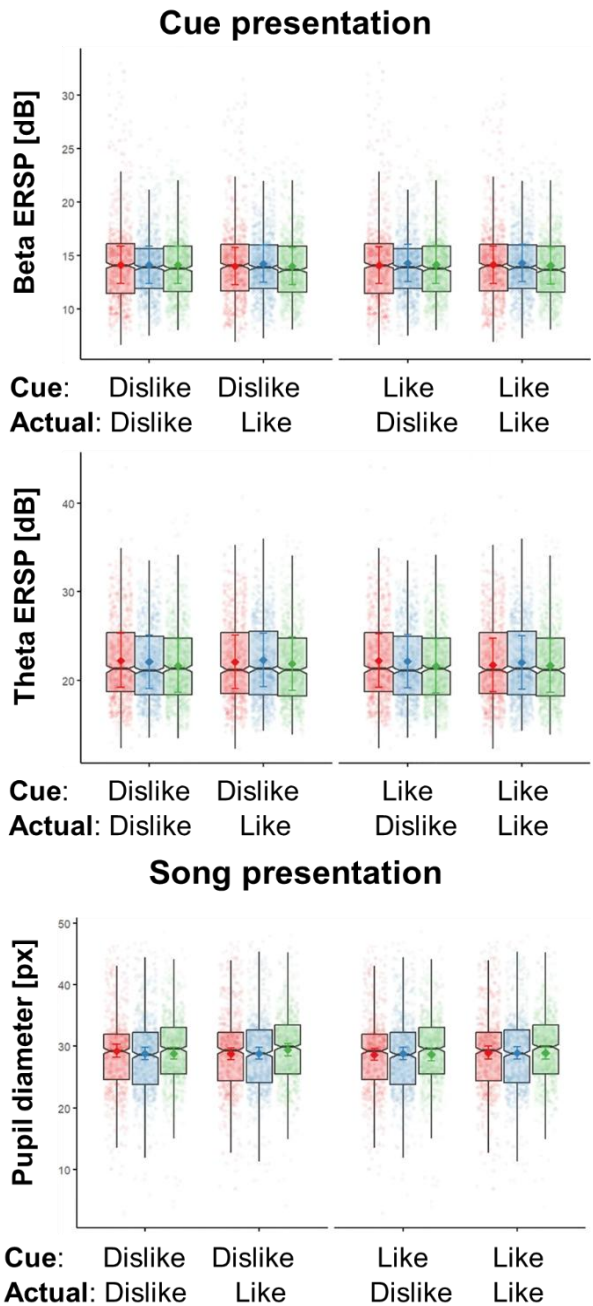

Figure 6: Effects of system accuracy, cue, and actual preference on mean oscillatory neural activity and pupil diameter during cue and music presentation. In the cue phase, significant differences in system accuracy were detected from subjects' event-related spectral perturbations (ERSP) in the beta and theta bands. In the music phase, resolving the significant three-way interaction revealed that pupil diameter decreased when a liked song was played following a dislike cue for the 80%-Accurate system, but increased for the 20%-Accurate system.

*Predicting reward from neurophysiological data*

We next tested the relationship between subjects' trial-wise song rating (Supplementary Tables 3 and 4), as well as expected reward (Supplementary Tables 5 and 6) and prediction error (Supplementary Tables 7 and 8) as derived from our computational model, and their ERSP and pupil diameter in the cue and music phases.

For song rating, we identified significant linear effects of ERSP in the alpha and beta band during music presentation (Figure 7). In both cases, increases in ERSP were related to decreases in rating, with no significant differences across systems (alpha: mean slope = –0.0725, –0.0457, –0.0783; beta: mean slope = –0.0614, –0.0607, –0.0742 for the 80%-, 50%-, 20%-accurate system, respectively). A quadratic effect of ERSP in the theta band was also detected in the music phase, although the trend was no longer significant after correcting for multiple comparisons. On the other hand, we identified a significant linear effect of pupil diameter in the cue phase (Figure 8), with increased dilation related to higher song ratings without significant differences across systems (mean slope = 0.0340, 0.0378, 0.0361 for 80%-, 50%-, 20%-accurate system, respectively). No ERSP effects in the cue phase and pupil effects in the music phase were detected.

For expected reward, the only significant neurophysiological effect was an accuracy*ERSP interaction in the alpha band in the music phase. While the quadratic model provided a better out-of-sample fit compared to the linear, the interaction was driven by a significant negative linear relationship between ERSP and expected reward specifically for the 80%-accurate system (mean slope = –0.0188, Z = –2.63, $p_{corrected}$ = 0.0252).

For prediction error, we found a significant main effect of ERSP in the alpha band in the music phase. Again, while the quadratic model was selected, the relationship was driven by a significant negative linear relationship between ERSP and prediction error, without significant differences in slope across systems (mean slope = –0.0383, –0.0305, –0.0478 for the 80%-, 50%-, 20%-accurate system, respectively). No other ERSP effects or pupil effects were detected in the cue and music phases.

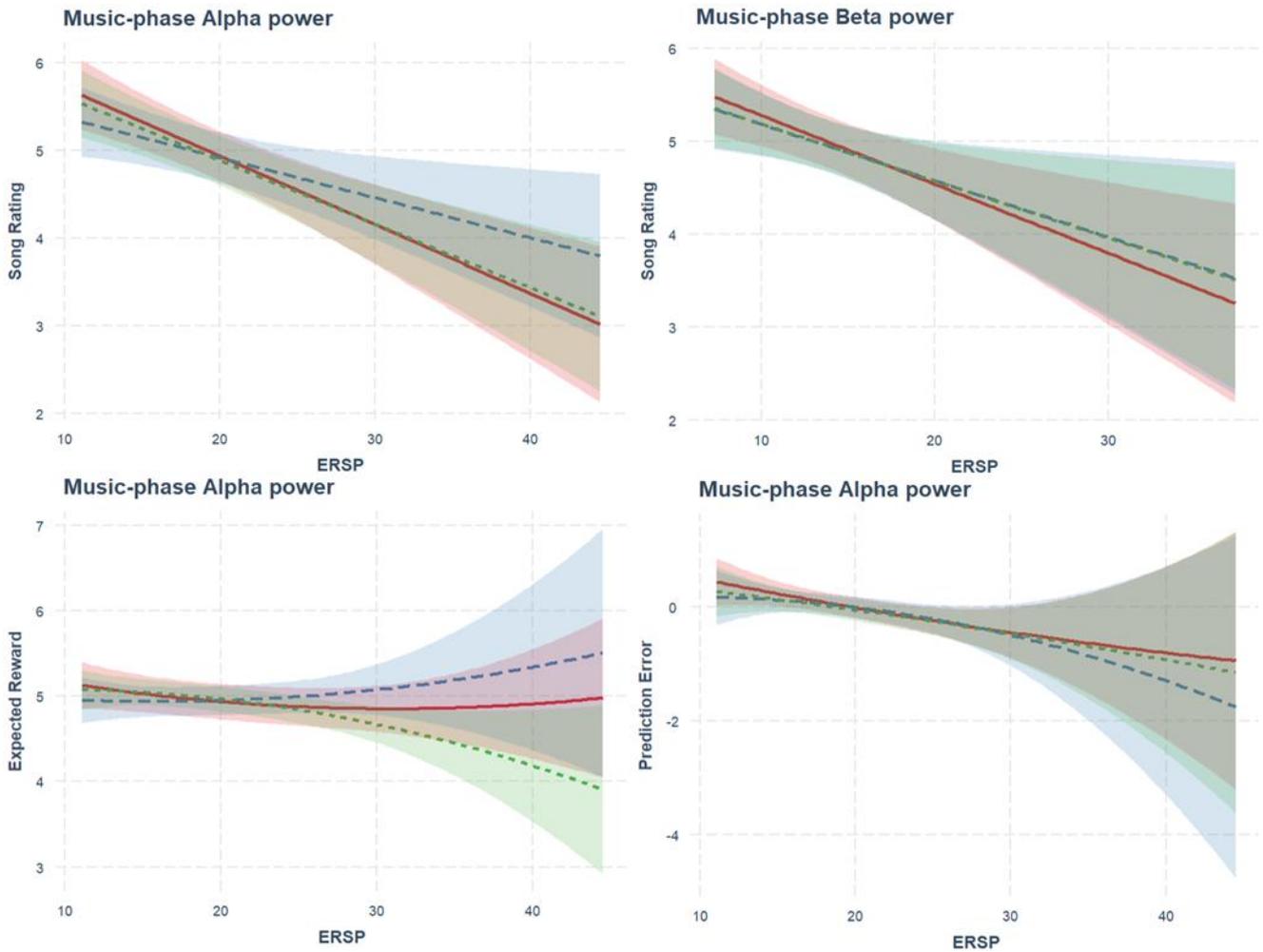

Figure 7. Oscillatory neural activity during music presentation—particularly in the alpha band—was significantly related to song rating, expected reward, and prediction error.

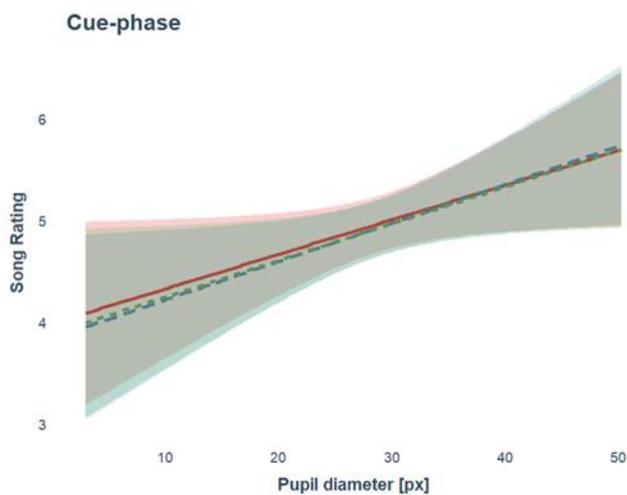

Figure 8. Pupil diameter during cue presentation was predictive of subsequent song rating, suggesting that pre-stimulus mental states may influence perception.

# Discussion

The current study sought to investigate the extent trust in automated systems shape their influence on user behaviour, as well as the underlying neurophysiological correlates. Taking music recommendation as a model system, we found that users rapidly established trust in the system that was calibrated to accuracy. In addition, positive and negative recommendation cues respectively modified users' song preference ratings up and down, for which the latter was modulated by users' trust. Modelling users' expected reward with a reinforcement learning model further revealed a reversal in the divergence between positive and negative expected reward when system recommendations were in opposition to users' preferences with high accuracy. With respect to neurophysiology, we found that differences in system accuracy were reflected in oscillatory activity in the beta and theta band, whilst alpha band activity was linearly related to song preference, prediction error, as well as expected reward when the system was trusted. An interaction with system accuracy and negative cue mismatches was also observed for pupil diameter. These show that automated systems can shape user behaviour—particularly when they are trustable, and that users' trust is dependent on neurocognitive mechanisms supporting expectation and monitoring.

Our results are particularly novel because they show that trusted external cues influence users' preferences and reward expectations. This is distinct from existing studies examining trust in automation as they rely on outcomes that have an objective truth assignment (such as detecting the presence of an object [14,16,17,38] or correct stimulus properties [11,12]. In those studies, trust is inferred by measuring differences to the extent the user deploys attention and resources towards monitoring the outputs of automation. By contrast, we show that automated systems directly modify users' subjective value judgments and expectations, and that is shaped by trust. This is in line with a previous study showing the influence of rating scale and system predictions on user opinions [39].

One crucial distinction between the current study and other error-monitoring or reward prediction paradigms is our use of music as the target rewarding stimulus. Although we were able to assign a ground truth on users' preference to a song based on their previous ratings, it is not clear at which point during music presentation that subjects assigned a preference value to the stimulus that was sufficiently stable to be carried over to the song rating phase. This is despite music being a well-known rewarding stimulus that engages the same neural pathways related to other rewards such as food, money, and sex [40–43]. It was therefore not possible for us to compare differences in neural activity in the time domain via typical event-related potentials (ERPs)—for which the oPE and oERN [10], FRN [9], or MMN [44] are examples of, as the precise alignment between stimuli could not be identified. This was our motivation to instead examine time-averaged changes across conditions in the frequency domain via ERSP. While unexpected or mismatched auditory events have been shown to also evoke significant differences in ERSP [45], its absence in the current study could indicate that those effects reflect unexpected differences in the auditory content of the stimulus itself and not the changes in reward associated with the stimulus.

On the other hand, differences in midfrontal theta oscillatory power across the three recommendation systems are in line with its proposed role in cognitive control [46,47]. While midfrontal theta has been associated with monitoring-related ERPs, such as novelty- or conflict-evoked N2, FRN, and ERN, it is important to note that midfrontal theta is not a direct counterpart of these time-domain signals but rather reflect gain adjustment to facilitate information processing across cortical areas and stimulus perception [48,49]. This could explain why we observed significantly larger theta activity in the 50%-accurate system—which has a higher uncertainty—compared to the 80%-accurate system, and that theta power differences were present both during cue and music presentation. Furthermore, as theta activity related to task monitoring is thought to be short-lasting [47], our averaging of ERSPs over time could explain why we did not detect any significant differences in oscillatory activity to cue and actual music preference mismatches.

In addition, the significant relationship between alpha power and song rating, expected reward (only in the 80%-accurate system), and prediction error suggests that alpha oscillatory activity represents general task-related processing as opposed to reward encoding. That is because we would have seen alpha power to relate specifically to song rating or prediction error if the latter were true. The proposed role of alpha power in task-related processing is consistent with the view that increased alpha desynchronisation (hence reduced oscillatory power) is related to increased attention via suppression of task-irrelevant information [50,51] that is likely modulated by thalamic activity [52].

The significant linear relationship between pupil size in the cue phase and subsequent song rating was unexpected. While reward expectations are thought to play a key role in musical pleasure [42,43,53], the lack of a significant difference across system accuracy suggests that the observed relationship was not due to differences in expected reward. This is corroborated by the absence of a significant relationship between pupil size and expected reward as predicted by the reinforcement learning model. One possible explanation could be that pupil size during cue presentation encodes pre-stimulus mental states that affect perception of subsequent stimulus. Indeed, a recent fMRI study has found that pre-stimulus functional connectivity between the auditory cortex and the mesolimbic reward network predicted psychophysiological responses to musical pleasure [54]. Relatedly, pupil diameter at cue could encode pre-stimulus arousal that modulates the emotional gain of the subsequent musical stimulus. This is supported by theories such as the arousal-biased competition model [55], which argue that arousal enhances processing and memory of salient stimuli via attention optimisation and activation of the locus coeruleus-norepinephrine (LC-NE) system [56].

Relatedly, the finding that pupil size varied as a function of system accuracy, cue, and actual music preference during stimulus presentation warrants discussion. Specifically, we found an increase in pupil diameter to mismatches given a dislike cue in the 80%- accurate system, but a decrease in the 20%-accurate system. While it may be difficult to interpret at first sight, this finding can be made sense of by noting that users implicitly regarded dislike cues as positive in the 20%-accurate system (Figure 4). As such, a mismatch in the 20%-system is not a dislike cue followed by an actual liked song, but rather a dislike cue followed by an actual disliked song. Therefore, this finding suggests that reward cue mismatches induce pupil dilation only when the cue is informative, consistent with its aforementioned role in arousal-based attention and studies examining the processing of auditory deviants (see [57] for review).

This study has some limitations. First, users' trust in automation was exclusively manipulated via the recommendation systems' accuracy. Hancock and colleagues [2] categorised additional factors that could affect trust: Human-related factors including personality traits, demographics, expertise, and current cognitive load [58–60], robot-related factors including competence (which we manipulated) [61,62], intent [63], extent of anthropomorphism [64] and system transparency [65–67], as well as environmental-related factors such as culture [68–71] and the context of interaction [72,73]. How these factors integrate towards a wholistic concept of trust in automation remains to be clarified.

Second, while our regression analysis allowed us to identify specific neurophysiological correlates of trust by observing differences across conditions, inference was made by pooling information across all trials from all subjects. This encoding approach allowed us to infer the underlying neurocognitive mechanisms but is in contrast to a decoding approach, where the aim is to classify trust based on an arbitrary combination of features on a trial-by-trial level [74,75]. Apart from developing decoding models that directly detect trust from behavioural and neurophysiological data in specific tasks, another exciting future application could be to enhance general computational models of trust (e.g., [76]) by providing dynamic feedback on users' mental states when interacting with automated systems [77] or consider using fMRI to enhance the spatial specificity of the current findings. Such efforts would prove promising towards the development of closed-loop trustable AI systems.

## Methods

*Subject Information*

We collected data from 30 healthy adults (15 females, age: M = 24.5 y, SD = 3.06 y). Subjects reported having normal hearing, no known history of neuropsychological disorders, severe cardiovascular diseases, liver and renal dysfunction, had no acute fever, and were not pregnant at the time of the experiment. Subjects who normally wore glasses either took them off or wore contact lenses during the experiment. Sample size was determined based on similar studies using EEG to detect auditory deviants [23–25]. Data from two females were excluded from analyses as their overall music reward scored below the preregistered cut-off 65 in the Barcelona Music Reward Questionnaire [26], which suggested potential musical anhedonia.

All subjects gave written informed consent prior to the experiment. Our study protocol was approved by the relevant competent IRB (name redacted for anonymisation) in accordance to the Declaration of Helsinki.

*Stimuli*

In a previous (unpublished) experiment, subjects were presented with the first 30s of the chorus from 64 songs sampled in a music streaming platform (name redacted for anonymisation) and rated how much they liked each excerpt on a 1 ('not at all') to 9 ('very much') scale. The songs were selected from over 500,000 songs on the platform to evenly span a 512-dimensional audio embedding space through self-supervised learning, ensuring sufficient variety across the selected songs.

For each subject, the first and last 10s of their top 15 rated songs were retained as a ground truth set of their liked musical stimuli. Similarly, the first and last 10s of their lowest 15 rated songs were retained as ground truth set of their disliked musical stimuli. This ground truth set of 30 liked and 30 disliked excerpts formed the individualised set of musical stimuli presented in the current experiment.

To minimise stimulus familiarity, the current experiment took place around 3 months after the initial stimulus rating session.

*Procedure*

The experiment (refer to Figure 1) consisted of three blocks, whose presentation order was counterbalanced across subjects. Each block consisted of 60 trials and represented song recommendations by one of three recommendation systems. Each trial began with a 4s-fixation cross, followed by a 4s-recommendation cue (the cue phase), and then 10s of music presentation (the music phase). The recommendation cue consisted of either the text 'You will like the upcoming song' (like cue) or 'You will dislike the upcoming song' (dislike cue). While like and dislike cues were displayed equally often, their accuracy varied across systems: Recommendation cues were accurate with 80%, 50%, or 20% probability, with correct and incorrect cues evenly assigned across like and dislike cues. The presented stimuli were randomly sampled from each subject's ground truth set of liked and disliked excerpts. To ensure consistency throughout the block, cues were pseudorandomised such that the distribution of correct and incorrect cues matched the assigned accuracy every 10 trials, and the same recommendation cue did not repeat more than twice in a row.

After song presentation, subjects rated how much they liked the musical excerpt on a 1 ('not at all') to 9 ('very much') scale using a mouse in a self-paced manner. At every 10$^{th}$ trial, subjects additionally rated how much they trusted the recommendation system in the same manner.

Musical stimuli were delivered at a comfortable volume via in-ear monitor earphones (Sony MDR-EX800ST). Experimental texts were presented in white against a grey background on a 27-inch LCD monitor (Philips 273VE5L) placed approximately 70 cm in front of the subject in a room lit at 280 lux. During the experiment, we also recorded subjects' brain activity via electroencephalography (EEG) and pupil size via eye-tracking. EEG data were recorded using an active 64-channel system (Brain Products GmbH) with Ag-Cl electrodes placed on the scalp in a 10-20 montage, whilst pupil sizes were recorded using Pupil Core (Pupil Labs GmbH). Both signals were sampled at 1 kHz.

*Modelling reward expectation*

We used a reinforcement learning model (the Rescorla-Wagner model [27]) to simulate subjects' learning of the association between seeing like or dislike cues and the eventual reward (music). Specifically, the expected reward $V$ for cue $i = \{like, dislike\}$ at the cue phase of trial $t + 1$ is given by $V_i(t + 1) = V_i(t) + \mathbf{1}_{cue(t)} a \Delta(t)$, where $a$ denotes the learning rate, $\Delta(t)$ the prediction error, and $\mathbf{1}_{cue(t)}$ is an indicator function equal to one if the presented cue at trial $t$ is $i$ and zero otherwise. The prediction error is given by $\Delta(t) = R(t) - V(t)$, where $R(t)$ is the reward received at trial $t$ (i.e., subjects' rating after hearing the stimuli in the music phase) and $V(t) = \Sigma_i \mathbf{1}_{cue(t)} V_i(t)$ is the expected reward of the presented cue. Here, we initialised the expected reward for both cues at trial 1 to be the mid-value of 5 on the $1 - 9$ scale.

While the learning rate $a$ is a constant in the original Rescorla-Wagner model, here we introduced a Pearce-Hall modification [28] to $a$ to better reflect the dependence of the learning rate on the magnitude of the prediction error [29]. In particular, $a$ is now dynamic, such that $a(t + 1) = b|\Delta(t)/8| + (1 - b)a(t)$ for some scaling parameter $b \in [0,1]$. Note that the prediction error is normalised to [0,1] by dividing it with the maximum possible prediction error (i.e., $9 - 1 = 8$).

Parameters for $a(1)$ and $b$ were estimated via grid search (between 0 and 1 in steps of 0.01) to minimise the total squared prediction error across all trials and blocks for each subject separately. This provides a maximum likelihood estimate for the two parameters, assuming that they were taken from a normal distribution.

*EEG data preprocessing*
EEG data were preprocessed using EEGLAB [30]. Recorded data were first resampled to 250 Hz, high-pass filtered with a 1 Hz cut-off, followed by powerline noise removal. Bad channels were then identified based on the local outlier factor algorithm [31] and replaced by interpolation via nearest neighbour average. A common average referencing was then carried out, followed by independent component analysis (ICA) to identify eye, muscle, heart, linear noise, or channel noise sources. Components identified with at least a 70% chance of containing those sources were subsequently removed.

Next, to derive the event-related spectral perturbation (ERSP), a wavelet decomposition (3 – 10 cycles) was applied at a [-2, 15] s time-window from cue onset at each trial within the 4 – 50 Hz frequency range. The mean power for each frequency band (theta: 4 – 8 Hz; alpha: 8 – 13 Hz; beta: 13 – 30 Hz) was then computed over the midfrontal electrodes F3, Fz, F4 for the cue phase and music phase for each trial, as well as the 1s baseline period preceding each phase. These were then mean-averaged across electrodes.

ERSP data were rejected from subsequent analyses if the baseline-differenced mean power at each phase exceeded 2.5 times the median absolute deviation (MAD) of the median for each frequency band in each subject. In total, 5.83% and 5.46% of trials in the cue and music phases were respectively rejected.

To ensure normality of the data, ERSP data at baseline and each frequency band were $10\log_{10}$-transformed.

*Pupil data preprocessing*
Pupil data were preprocessed by first extracting the diameter of each pupil using a 3D eye model from the eye tracker's native software. Motion and eyeblink artefacts were identified and censored whenever they exceeded a confidence threshold of 0.5 and filter threshold of 200 ms. Data were further censored if the estimated pupil diameter erroneously fell below 0. Pupil diameter for both eyes were then mean averaged and epoched to the cue and music phases of each trial with a 200 ms baseline included. Epochs with over 40% artefacts were removed. The remaining epochs were down-sampled to 10 Hz to conserve disk space, and the censored data were replaced by linear interpolation. The pupil diameter during each phase, as well as its corresponding baseline period, were then mean-averaged over time.

Pupil data were rejected from subsequent analyses if the baseline-differenced diameter of each phase exceeded 2.5 MAD of the median in each subject. In total, 5.91% and 4.64% of trials in the cue and music phases were respectively rejected.

*Statistical analyses*

We fitted linear mixed-effect models to quantify the statistical relationship between variables. Linear mixed-effect models have the advantage of information partial pooling as they incorporate trial-wise data hierarchically nested within subjects and stimuli, as opposed to traditional ANOVA where trials are first aggregated within subject [32]. Here, we treat subject and excerpt as two crossed-random effect groupings. When predicting ERSP and pupil diameter, trial-wise baselines were explicitly added as predictors to improve precision and power, as opposed to using baseline-differenced values [33,34]. To strike a balance between maintaining power without inflating Type I error rate [35], we further selected the optimal random effects structure, derived by stepwise elimination from the maximal model, using Akaike's Information Criterion (AIC). Models were estimated via maximum likelihood with Satterthwaite's approximation for degrees of freedom.

We also fitted models that sought to predict song preference, expected reward, and prediction error from ERSP or pupil diameter. We considered quadratic effects as increasing studies have demonstrated similar neurophysiological responses to both positive and negative reward stimuli [36,37], suggesting that linear models may be inadequate. For the quadratic model, fixed effect predictors included an interaction between system accuracy and orthogonalised quadratic effect of ERSP or pupil diameter, an interaction between system accuracy and orthogonalised linear effect of ERSP or pupil diameter, the trial-wise baseline ERSP or pupil diameter, as well as their lower-order terms. For the linear model, the same predictors were included without the quadratic term and without orthogonalisation. Model order was selected using corrected AIC to account for better model fit due to increased number of parameters. If the difference in criteria was less than 2, the lower-order model was selected for parsimony.

We used null-hypothesis testing for statistical inference and we report results from two-tailed tests using an alpha-threshold of 0.05. Pairwise comparisons of empirical marginal means for post hoc tests were corrected for multiple comparisons via Tukey's method.

76. Williams, K.J., Yuh, M.S., and Jain, N. (2023). A Computational Model of Coupled Human Trust and Self-confidence Dynamics. ACM Trans. Human-Robot Interact. *12*, 1–29. Available at: https://dl.acm.org/doi/10.1145/3594715.

77. Rodriguez Rodriguez, L., Bustamante Orellana, C.E., Chiou, E.K., Huang, L., Cooke, N., and Kang, Y. (2023). A review of mathematical models of human trust in automation. Front. Neuroergonomics *4*. Available at: https://www.frontiersin.org/articles/10.3389/fnrgo.2023.1171403/full.
24

# Supplementary Materials
**Cue phase**

*Alpha-ERSP*

|  | $df_{numerator}$ | $df_{denominator}$ | F | p |
|---|---|---|---|---|
| Accuracy | 2 | 4477 | 0.166 | 0.847 |
| Cue | 1 | 4609 | 0.292 | 0.589 |
| Actual | 1 | 203 | 0.0364 | 0.849 |
| Baseline | 1 | 4584 | 5141 | $<2.00 \times 10^{-16}$ *** |
| Accuracy*Cue | 2 | 4614 | 0.00270 | 0.997 |
| Accuracy*Actual | 2 | 4477 | 0.847 | 0.429 |
| Cue*Actual | 1 | 4608 | 0.166 | 0.684 |
| Accuracy*Cue*Actual | 2 | 4614 | 0.189 | 0.828 |

*Beta-ERSP*

|  | $df_{numerator}$ | $df_{denominator}$ | F | p |
|---|---|---|---|---|
| Accuracy | 2 | 4543 | 3.99 | 0.0185 * |
| Cue | 1 | 4506 | 1.86 | 0.173 |
| Actual | 1 | 4501 | 0.356 | 0.551 |
| Baseline | 1 | 26.7 | 147 | $2.72 \times 10^{-12}$ *** |
| Accuracy*Cue | 2 | 4508 | 0.0956 | 0.909 |
| Accuracy*Actual | 2 | 4509 | 1.01 | 0.365 |
| Cue*Actual | 1 | 4508 | 0.208 | 0.649 |
| Accuracy*Cue*Actual | 2 | 4508 | 0.989 | 0.372 |

*Theta-ERSP*

|  | $df_{numerator}$ | $df_{denominator}$ | F | p |
|---|---|---|---|---|
| Accuracy | 2 | 4603 | 10.5 | $2.89 \times 10^{-5}$ *** |
| Cue | 1 | 4564 | 3.25 | 0.0714 |
| Actual | 1 | 4567 | 0.442 | 0.506 |
| Baseline | 1 | 26.2 | 59.6 | $3.29 \times 10^{-8}$ *** |
| Accuracy*Cue | 2 | 4566 | 0.0919 | 0.912 |
| Accuracy*Actual | 2 | 4566 | 2.58 | 0.0758 |
| Cue*Actual | 1 | 4566 | 2.45 | 0.118 |
| Accuracy*Cue*Actual | 2 | 4566 | 0.0414 | 0.959 |

*Pupil diameter*

|  | $df_{numerator}$ | $df_{denominator}$ | F | p |
|---|---|---|---|---|
| Accuracy | 2 | 26.8 | 0.684 | 0.513 |
| Cue | 1 | 3907 | 1.48 | 0.224 |
| Actual | 1 | 3908 | 0.0865 | 0.769 |
| Baseline | 1 | 3489 | 9289 | $<2.00 \times 10^{-16}$ *** |
| Accuracy*Cue | 2 | 3908 | 0.0956 | 0.909 |
| Accuracy*Actual | 2 | 3909 | 2.18 | 0.113 |
| Cue*Actual | 1 | 3908 | 1.78 | 0.182 |
| Accuracy*Cue*Actual | 2 | 3908 | 0.436 | 0.646 |

Supplementary Table 1. Summary table of the effects of system accuracy (80%/50%/20%), recommendation cue (like/dislike), actual preference (like/dislike) on neural oscillatory activity and pupil diameter in the cue phase. *: $p < 0.05$, **: = $p < 0.01$, ***: $p < 0.001$.

**Music phase**

*Alpha-ERSP*

|  | $df_{numerator}$ | $df_{denominator}$ | F | p |
|---|---|---|---|---|
| Accuracy | 2 | 27.2 | 0.367 | 0.697 |
| Cue | 1 | 4590 | 0.347 | 0.556 |
| Actual | 1 | 4219 | 3.97 | 0.0465 * |
| Baseline | 1 | 4661 | 758 | $<2.00 \times 10^{-16}$ *** |
| Accuracy*Cue | 2 | 4592 | 0.621 | 0.537 |
| Accuracy*Actual | 2 | 4520 | 0.403 | 0.668 |
| Cue*Actual | 1 | 4595 | 0.137 | 0.711 |
| Accuracy*Cue*Actual | 2 | 4593 | 0.166 | 0.847 |

*Beta-ERSP*

|  | $df_{numerator}$ | $df_{denominator}$ | F | p |
|---|---|---|---|---|
| Accuracy | 2 | 26.8 | 0.402 | 0.673 |
| Cue | 1 | 4500 | 0.891 | 0.345 |
| Actual | 1 | 3992 | 3.47 | 0.0627 |
| Baseline | 1 | 4551 | 2765 | $<2.00 \times 10^{-16}$ *** |
| Accuracy*Cue | 2 | 4499 | 0.304 | 0.738 |
| Accuracy*Actual | 2 | 4433 | 1.95 | 0.142 |
| Cue*Actual | 1 | 4501 | $2.00 \times 10^{-4}$ | 0.989 |
| Accuracy*Cue*Actual | 2 | 4501 | 0.974 | 0.378 |

*Theta-ERSP*

|  | $df_{numerator}$ | $df_{denominator}$ | F | p |
|---|---|---|---|---|
| Accuracy | 2 | 4577 | 6.64 | 0.00132 ** |
| Cue | 1 | 4551 | 0.0686 | 0.793 |
| Actual | 1 | 4552 | 2.13 | 0.145 |
| Baseline | 1 | 26.7 | 51.3 | $1.14 \times 10^{-7}$ *** |
| Accuracy*Cue | 2 | 4551 | 1.03 | 0.357 |
| Accuracy*Actual | 2 | 4550 | 0.0823 | 0.921 |
| Cue*Actual | 1 | 4551 | 0.105 | 0.746 |
| Accuracy*Cue*Actual | 2 | 4551 | 0.274 | 0.760 |

*Pupil diameter*

|  | $df_{numerator}$ | $df_{denominator}$ | F | p |
|---|---|---|---|---|
| Accuracy | 2 | 3949 | 0.370 | 0.691 |
| Cue | 1 | 3976 | 3.30 | 0.0670 |
| Actual | 1 | 3738 | 2.03 | 0.155 |
| Baseline | 1 | 25.6 | 371 | $<2.00 \times 10^{-16}$ *** |
| Accuracy*Cue | 2 | 3987 | 1.83 | 0.160 |
| Accuracy*Actual | 2 | 3924 | 2.25 | 0.106 |
| Cue*Actual | 1 | 3988 | 0.976 | 0.323 |
| Accuracy*Cue*Actual | 2 | 3987 | 3.09 | 0.0456 * |

Supplementary Table 2. Summary table of the effects of system accuracy (80%/50%/20%), recommendation cue (like/dislike), actual preference (like/dislike) on neural oscillatory activity and pupil diameter in the music phase. *: p < 0.05, **: = p < 0.01, ***: p < 0.001.

**Song rating – Cue phase**

*Alpha-ERSP (linear, dAICc = 4.8)*

|  | $df_{numerator}$ | $df_{denominator}$ | F | p |
|---|---|---|---|---|
| Accuracy | 2 | 4511 | 2.10 | 0.123 |
| ERSP | 1 | 3420 | 0.0204 | 0.887 |
| Baseline | 1 | 4222 | 1.52 | 0.218 |
| Accuracy*ERSP | 2 | 4513 | 1.95 | 0.142 |

*Beta-ERSP (quadratic, dAICc = 39.6)*

|  | $df_{numerator}$ | $df_{denominator}$ | F | p |
|---|---|---|---|---|
| Accuracy | 2 | 4343 | 0.133 | 0.875 |
| ERSP | 2 | 1163 | 2.10 | 0.122 |
| Baseline | 1 | 4394 | 2.98 | 0.0843 |
| Accuracy*ERSP | 4 | 3384 | 1.96 | 0.0979 |

*Theta-ERSP (quadratic, dAICc = 30.2)*

|  | $df_{numerator}$ | $df_{denominator}$ | F | p |
|---|---|---|---|---|
| Accuracy | 2 | 4360 | 0.0581 | 0.944 |
| ERSP | 2 | 185 | 0.131 | 0.878 |
| Baseline | 1 | 4320 | 0.0199 | 0.888 |
| Accuracy*ERSP | 4 | 2086 | 0.963 | 0.427 |

*Pupil diameter (linear, dAICc = 4.3)*

|  | $df_{numerator}$ | $df_{denominator}$ | F | p |
|---|---|---|---|---|
| Accuracy | 2 | 3884 | 0.0874 | 0.916 |
| Diameter | 1 | 3817 | 5.40 | 0.0202 * |
| Baseline | 1 | 3911 | 0.813 | 0.367 |
| Accuracy*Diameter | 2 | 3883 | 0.0482 | 0.953 |

Supplementary Table 3. Summary table on predicting song rating from system accuracy (80%/50%/20%), neural oscillatory activity, and pupil diameter in the cue phase. Quadratic and linear models were considered and selected based on relative difference in corrected Akaike's Information Criterion (dAICc). *: p < 0.05, **: = p < 0.01, ***: p < 0.001.

**Song rating – Music phase**

*Alpha-ERSP (linear, dAICc = 2.3)*

|  | $df_{numerator}$ | $df_{denominator}$ | F | p |
|---|---|---|---|---|
| Accuracy | 2 | 4565 | 1.93 | 0.145 |
| ERSP | 1 | 3065 | 22.9 | $1.83 \times 10^{-6}$ *** |
| Baseline | 1 | 4588 | 5.75 | 0.0165 * |
| Accuracy*ERSP | 2 | 4566 | 1.82 | 0.161 |

*Beta-ERSP (linear, dAICc = 1.7)*

|  | $df_{numerator}$ | $df_{denominator}$ | F | p |
|---|---|---|---|---|
| Accuracy | 2 | 4466 | 0.380 | 0.684 |
| ERSP | 1 | 4215 | 9.48 | 0.00209 ** |
| Baseline | 1 | 4508 | 3.39 | 0.0657 |
| Accuracy*ERSP | 2 | 4462 | 0.265 | 0.767 |

*Theta-ERSP (quadratic, dAICc = 17.1)*

|  | $df_{numerator}$ | $df_{denominator}$ | F | p |
|---|---|---|---|---|
| Accuracy | 2 | 4457 | 0.132 | 0.876 |
| ERSP | 2 | 2596 | 1.09 | 0.337 |
| Baseline | 1 | 4513 | 0.794 | 0.373 |
| Accuracy*ERSP | 4 | 4487 | 3.23 | 0.0118 * |

*Pupil diameter (quadratic, dAICc = 21.9)*

|  | $df_{numerator}$ | $df_{denominator}$ | F | p |
|---|---|---|---|---|
| Accuracy | 2 | 2655 | 0.484 | 0.616 |
| Diameter | 2 | 20.4 | 2.13 | 0.144 |
| Baseline | 1 | 3914 | 0.823 | 0.364 |
| Accuracy*Diameter | 4 | 1759 | 0.847 | 0.495 |

Supplementary Table 4. Summary table on predicting song rating from system accuracy (80%/50%/20%), neural oscillatory activity, and pupil diameter in the music phase. Quadratic and linear models were considered and selected based on relative difference in corrected Akaike's Information Criterion (dAICc). *: $p < 0.05$, **: = $p < 0.01$, ***: $p < 0.001$.

**Expected reward – Cue phase**

*Alpha-ERSP (quadratic, dAICc = 38.4)*

|  | $df_{numerator}$ | $df_{denominator}$ | F | p |
|---|---|---|---|---|
| Accuracy | 2 | 25.5 | 0.213 | 0.810 |
| ERSP | 2 | 4246 | 1,41 | 0.245 |
| Baseline | 1 | 4578 | 0.278 | 0.598 |
| Accuracy*ERSP | 4 | 356 | 1.19 | 0.314 |

*Beta-ERSP (quadratic, dAICc = 39.8)*

|  | $df_{numerator}$ | $df_{denominator}$ | F | p |
|---|---|---|---|---|
| Accuracy | 2 | 25.7 | 0.248 | 0.782 |
| ERSP | 2 | 3171 | 2.31 | 0.0990 |
| Baseline | 1 | 4553 | 0.158 | 0.691 |
| Accuracy*ERSP | 4 | 292 | 1.50 | 0.201 |

*Theta-ERSP (quadratic, dAICc = 39.1)*

|  | $df_{numerator}$ | $df_{denominator}$ | F | p |
|---|---|---|---|---|
| Accuracy | 2 | 25.0 | 0.150 | 0.862 |
| ERSP | 2 | 3418 | 0.425 | 0.654 |
| Baseline | 1 | 4477 | 0.249 | 0.618 |
| Accuracy*ERSP | 4 | 383 | 1.67 | 0.157 |

*Pupil diameter (quadratic, dAICc = 41.6)*

|  | $df_{numerator}$ | $df_{denominator}$ | F | p |
|---|---|---|---|---|
| Accuracy | 2 | 23.6 | 0.0540 | 0.948 |
| Diameter | 2 | 3858 | 0.940 | 0.391 |
| Baseline | 1 | 3960 | 3.56 | 0.0593 |
| Accuracy*Diameter | 4 | 324 | 1.47 | 0.210 |

Supplementary Table 5. Summary table on predicting expected reward during cue presentation from system accuracy (80%/50%/20%), neural oscillatory activity, and pupil diameter in the cue phase. Quadratic and linear models were considered and selected based on relative difference in corrected Akaike's Information Criterion (dAICc). *: $p < 0.05$, **: = $p < 0.01$, ***: $p < 0.001$.

**Expected reward – Music phase**

*Alpha-ERSP (quadratic, dAICc = 153.8)*

|  | $df_{numerator}$ | $df_{denominator}$ | F | p |
|---|---|---|---|---|
| Accuracy | 2 | 25.1 | 0.0407 | 0.960 |
| ERSP | 2 | 2885 | 2.08 | 0.125 |
| Baseline | 1 | 4668 | 0.236 | 0.628 |
| Accuracy*ERSP | 4 | 228 | 3.36 | 0.0107 * |

*Beta-ERSP (quadratic, dAICc = 40.0)*

|  | $df_{numerator}$ | $df_{denominator}$ | F | p |
|---|---|---|---|---|
| Accuracy | 2 | 25.3 | 0.203 | 0.817 |
| ERSP | 2 | 1703 | 0.808 | 0.446 |
| Baseline | 1 | 4573 | 2.12 | 0.146 |
| Accuracy*ERSP | 4 | 201 | 2.01 | 0.0951 |

*Theta-ERSP (quadratic, dAICc = 42.6)*

|  | $df_{numerator}$ | $df_{denominator}$ | F | p |
|---|---|---|---|---|
| Accuracy | 2 | 25.3 | 0.222 | 0.803 |
| ERSP | 2 | 1521 | 0.256 | 0.774 |
| Baseline | 1 | 4589 | 0.0195 | 0.889 |
| Accuracy*ERSP | 4 | 149 | 2.02 | 0.0946 |

*Pupil diameter (quadratic, dAICc = 38.6)*

|  | $df_{numerator}$ | $df_{denominator}$ | F | p |
|---|---|---|---|---|
| Accuracy | 2 | 23.9 | 0.0198 | 0.980 |
| Diameter | 2 | 2961 | 0.00841 | 0.992 |
| Baseline | 1 | 4016 | 2.02 | 0.156 |
| Accuracy*Diameter | 4 | 240 | 0.343 | 0.849 |

Supplementary Table 6. Summary table on predicting expected reward during cue presentation from system accuracy (80%/50%/20%), neural oscillatory activity, and pupil diameter in the music phase. Quadratic and linear models were considered and selected based on relative difference in corrected Akaike's Information Criterion (dAICc). *: $p < 0.05$, **: = $p < 0.01$, ***: $p < 0.001$.

**Prediction error – Cue phase**

*Alpha-ERSP (quadratic, dAICc = 2.2)*

|  | $df_{numerator}$ | $df_{denominator}$ | F | p |
|---|---|---|---|---|
| Accuracy | 2 | 4482 | 0.0955 | 0.909 |
| ERSP | 2 | 681 | 0.360 | 0.698 |
| Baseline | 1 | 1881 | 1.71 | 0.191 |
| Accuracy*ERSP | 4 | 4047 | 0.225 | 0.924 |

*Beta-ERSP (quadratic, dAICc = 7.4)*

|  | $df_{numerator}$ | $df_{denominator}$ | F | p |
|---|---|---|---|---|
| Accuracy | 2 | 4428 | 0.122 | 0.886 |
| ERSP | 2 | 738 | 1.01 | 0.365 |
| Baseline | 1 | 2919 | 3.06 | 0.0805 |
| Accuracy*ERSP | 4 | 2162 | 0.996 | 0.408 |

*Theta-ERSP (quadratic, dAICc = 42.1)*

|  | $df_{numerator}$ | $df_{denominator}$ | F | p |
|---|---|---|---|---|
| Accuracy | 2 | 4364 | 0.0741 | 0.929 |
| ERSP | 2 | 148 | 0.261 | 0.770 |
| Baseline | 1 | 2079 | 0.0133 | 0.908 |
| Accuracy*ERSP | 4 | 1971 | 0.679 | 0.607 |

*Pupil diameter (quadratic, dAICc = 30.6)*

|  | $df_{numerator}$ | $df_{denominator}$ | F | p |
|---|---|---|---|---|
| Accuracy | 2 | 3782 | 0.268 | 0.765 |
| Diameter | 2 | 153 | 2.05 | 0.132 |
| Baseline | 1 | 2997 | 2.34 | 0.126 |
| Accuracy*Diameter | 4 | 1241 | 0.0953 | 0.984 |

Supplementary Table 7. Summary table on predicting reward prediction errors from system accuracy (80%/50%/20%), neural oscillatory activity, and pupil diameter in the cue phase. Quadratic and linear models were considered and selected based on relative difference in corrected Akaike's Information Criterion (dAICc). *: p < 0.05, **: = p < 0.01, ***: p < 0.001.

**Prediction error – Music phase**

*Alpha-ERSP (quadratic, dAICc = 2.5)*

|  | $df_{numerator}$ | $df_{denominator}$ | F | p |
|---|---|---|---|---|
| Accuracy | 2 | 4535 | 0.416 | 0.660 |
| ERSP | 2 | 372 | 5.89 | 0.00302 ** |
| Baseline | 1 | 3636 | 9.36 | 0.00235 ** |
| Accuracy*ERSP | 4 | 4070 | 0.226 | 0.924 |

*Beta-ERSP (quadratic, dAICc = 43.5)*

|  | $df_{numerator}$ | $df_{denominator}$ | F | p |
|---|---|---|---|---|
| Accuracy | 2 | 4350 | 0.303 | 0.739 |
| ERSP | 2 | 294 | 2.79 | 0.0631 |
| Baseline | 1 | 4207 | 6.43 | 0.0112 * |
| Accuracy*ERSP | 4 | 680 | 0.510 | 0.729 |

*Theta-ERSP (quadratic, dAICc = 111.9)*

|  | $df_{numerator}$ | $df_{denominator}$ | F | p |
|---|---|---|---|---|
| Accuracy | 2 | 4296 | 0.383 | 0.682 |
| ERSP | 2 | 162 | 0.505 | 0.605 |
| Baseline | 1 | 3917 | 1.19 | 0.276 |
| Accuracy*ERSP | 4 | 3088 | 1.28 | 0.275 |

*Pupil diameter (linear, dAICc = 4.2)*

|  | $df_{numerator}$ | $df_{denominator}$ | F | P |
|---|---|---|---|---|
| Accuracy | 2 | 3624 | 0.488 | 0.614 |
| Diameter | 1 | 1437 | 2.68 | 0.102 |
| Baseline | 1 | 3728 | 0.00359 | 0.952 |
| Accuracy*Diameter | 2 | 3662 | 0.541 | 0.582 |

Supplementary Table 8. Summary table on predicting reward prediction errors from system accuracy (80%/50%/20%), neural oscillatory activity, and pupil diameter in the music phase. Quadratic and linear models were considered and selected based on relative difference in corrected Akaike's Information Criterion (dAICc). *: p < 0.05, **: = p < 0.01, ***: p < 0.001.